# Probing the Design Space of InSb Topological Superconductor Nanowires for the Realization of Majorana Zero Modes

M. Poljak

Computational Nanoelectronics Group
University of Zagreb Faculty of Electrical Engineering and Computing, Zagreb, Croatia
E-mail: mirko.poljak@fer.unizg.hr

*Abstract*—Non-Abelian anyons such as Majorana zero modes (MZMs) have the potential to enable fault-tolerant quantum computing through topological protection. Experimentally reported InSb topological superconductor nanowires (TSNW) are investigated theoretically and numerically to evaluate their suitability to host MZMs. We employ eigenspectra analysis and quantum transport based on the non-equilibrium Green's function (NEGF) formalism to investigate the eigenenergies, Majorana wave functions via local density of states, transmission spectra for Andreev processes, and zero-bias conductance peaks (ZBCPs) in InSb TSNWs. For 1.6 μm- and 2.2 μm-long InSb TSNWs we demonstrate the existence of the optimum design space defined by the applied magnetic field and electrochemical potential, which leads to clear ZBCP signatures with a Majorana localization length down to ~340 nm.

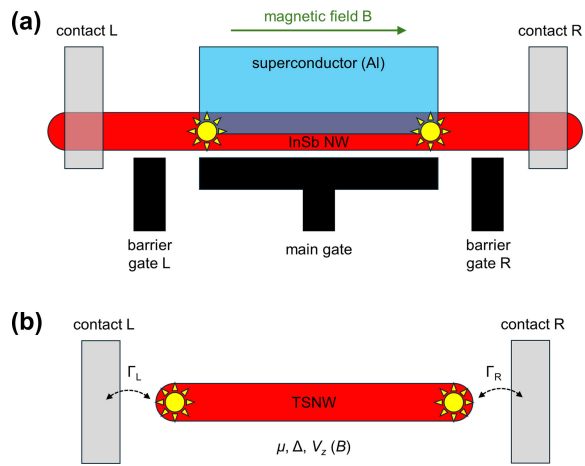

Fig. 1. (a) Illustration of the experimental setup of TSNW hybrid devices and (b) simplified system for NEGF simulations. Star-like symbols designate the Majorana modes.

## I. INTRODUCTION

Quantum computing has recently demonstrated computational advantage of quantum processing units (QPUs) over the classical counterparts [1], which moves it closer to real-world practical applications in areas such as cybersecurity, climate modeling, material prediction and drug discovery [2]. While there are many physical realizations of qubits in QPUs, the one based on exotic quasiparticles called non-Abelian anyons is very promising for fault-tolerant computation due to its innate immunity to local perturbations [3], [4]. Practical realizations of non-Abelian anyons include producing Majorana zero modes (MZMs) or Majorana bound states (MBSs) in topological superconductor nanowires (TSNWs) [5], [6] or in proximitized chains of quantum dots [7], [8]. Both approaches stem from a toy model proposed by Kitaev [9], who demonstrated the existence of MZMs at the ends of the so-called Kitaev chain. Evidence of MZMs was hinted at in [5] through the measured zero-bias conductance peaks in an InSb nanowire (NW) with proximitized superconductivity in a strong magnetic field, as illustrated in Fig. 1a. In addition, proof of passing the topological protocol has been recently reported for InAs-based Majorana hybrid devices [10] and, moreover, a device architecture compatible for future braiding experiments with millions of topological qubits has been introduced [11]. However, localization and braiding features have not been demonstrated yet, because the effects of finite temperature and disorder seem to be currently insurmountable [12]. Theoretical and computational work in the literature suggest that increasing the TSNW length is needed for topological protection [13], [14], which opens questions on the desirable design space and biasing conditions for realizing MZMs in TSNWs.

The TSNWs fabricated so far have various dimensions with the lengths ranging from only ~300 nm up to ~3 μm [5], [10], [15]. Additionally, biasing by local gate electrodes, magnetic field, and proximitized superconductivity vary significantly in the literature, and the optimum conditions for realizing technologically feasible MZMs are unknown. In this paper, we assess the design space for InSb TSNWs by employing the non-equilibrium Green's function (NEGF) formalism. The basic characteristics of MZMs are explored in terms of their dependence on the system design parameters such as nanowire length, electrochemical biasing, proximitized superconductivity and the magnetic field. In addition, we use NEGF to analyze the transport properties and conductance at 0 K that is mainly determined by the Andreev processes. Finally, we determine the optimum design space or biasing conditions for the 1.6 μm- and 2.2 μm-long InSb TSNWs, which could enable technologically relevant MZMs and contribute towards practical topological quantum computing (TQC).

## II. Theory and Methodology

Majorana quasiparticles are zero-energy modes localized at the edges of a 1D topological superconductor, according to the Lutchyn-Oreg model [16], [17] based on the original idea by Kitaev [9]. A schematic illustration of experimental devices is shown in Fig. 1a, while the simplified model for numerical simulations is illustrated in Fig. 1b. The tight-binding Hamiltonian of the TSNW is

$$\mathbf{H}_{TSNW} = \sum_i c_i^\dagger \left[ (2t - \mu)\tau_z \otimes \sigma_0 + V_z \tau_z \otimes \sigma_x + \right.$$
$$\left. + \Delta \tau_y \otimes \sigma_y \right] c_i + \quad (1)$$
$$+ \sum_{(i,j)} c_i^\dagger \left[ -t \tau_z \otimes \sigma_0 + i t_{SO} \tau_z \otimes \sigma_y \right] c_j ,$$

where $c_i = (c_{i\uparrow}, c_{i\downarrow}, c_{i\uparrow}^\dagger, c_{i\downarrow}^\dagger)$ is the Nambu spinor at site $i$ ($i = 1, …, N$), $\mu$ is the electrochemical potential, $t$ is the nearest neighbor hopping parameter, $\Delta$ is the superconductor pairing parameter or proximity-induced superconducting gap, $V_z$ is the Zeeman field due to the magnetic field $B$, i.e. $V_z = g\mu_B B/2$, and $t_{SO}$ is the Rashba spin-orbit coupling parameter that depends on the Rashba coupling strength $\alpha_R$, i.e. $t_{SO} = \alpha_R/(2a)$ [18], [19]. Within the single-band EMA model of the InSb TSNW, $a$ is the lattice spacing and the finite-difference hopping parameter becomes $t = \hbar^2/(2m^*a^2)$, where $m^*$ is the electron effective mass in the nanowire. In Eq. (1), matrices $\sigma$ are Pauli matrices in the spin basis, and matrices $\tau$ are Nambu matrices in the particle-hole space [20]. We assume that the main gate sets a constant electrochemical potential, and that the superconductor sets a constant superconducting gap in the entire device (see Fig. 1a). Any disorder is neglected, and the remaining parameters of the InSb TSNW Hamiltonian are taken from [18].

The transmission function and conductance of the InSb TSNW is calculated using our in-house NEGF code [14], [21], [22], [23], [24], which is based on the Keldysh formalism [25], [26], [27] and the work of Kadanoff and Baym [28]. The main point of the NEGF approach is to find the retarded Green's function of the device

$$\mathbf{G}^r(E) = \left[ (E + i0^+) \mathbf{I}_{4N} - \mathbf{H}_{TSNW} - \mathbf{\Sigma}_L^r - \mathbf{\Sigma}_R^r \right]^{-1} \quad (2)$$

where $\mathbf{\Sigma}$ represent the retarded contact self-energy matrices that account for open boundary conditions, i.e. interactions between the TSNW and the two normal electrodes. From the retarded Green's function, we calculate the transmission as

$$T(E) = Tr\left[ \mathbf{\Gamma}_L^r \mathbf{G}^r(E) \mathbf{\Gamma}_R^r \mathbf{G}^a(E) \right], \quad (3)$$

where $\mathbf{G}^a(E)$ is the advanced Green's function of the device, i.e. $\mathbf{G}^a(E) = [\mathbf{G}^r(E)]^\dagger$, and $\mathbf{\Gamma}$ are the broadening matrices [26] that are defined here within the wide-band limit (WBL) approximation [29], [30], [31], [32]. In the analysis of transport properties of InSb TSNWs we need to consider three mechanisms: direct tunneling, Andreev reflection (AR) and crossed Andreev reflection (CAR) [33], [34]. The individual transmission components are calculated as in [18], [34].

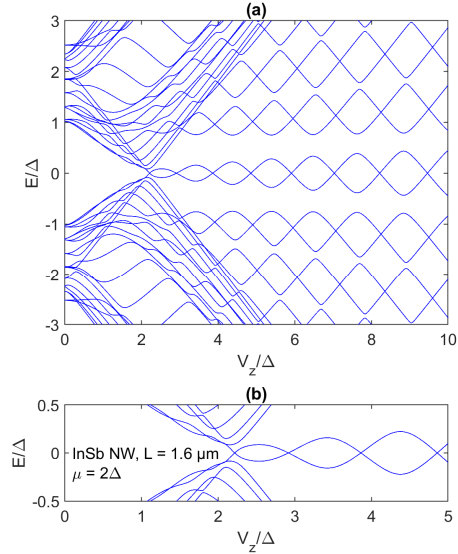

Fig. 2. Eigenspectrum of the 1.6 μm-long InSb NW vs. Zeeman field for the electrochemical potential set to 0.5 meV.

From the transmission using the Landauer formula we calculate the conductance, which is a real-world macroscopic observable that is accessible in experiments [5]. The conductance matrix for the two-probe ($L$ and $R$) device configuration equals

$$G = \begin{bmatrix} G_{LL} & G_{LR} \\ G_{RL} & G_{RR} \end{bmatrix} = \begin{bmatrix} \frac{\partial I_L}{\partial V_L}\big|_{V_R=0} & \frac{\partial I_L}{\partial V_R}\big|_{V_L=0} \\ \frac{\partial I_R}{\partial V_L}\big|_{V_R=0} & \frac{\partial I_R}{\partial V_R}\big|_{V_L=0} \end{bmatrix}, \quad (4)$$

where the diagonal components are the local conductances at the two contacts and the off-diagonal elements are the non-local conductances [35]. In our calculations, we set the contact broadening ($\Gamma_L$, $\Gamma_R$) to 0.1 eV, so that $\Sigma_{L,R}$ is −i0.05 eV in the WBL contacts. Depending on the conductance measurement setup, the TSNW can exhibit a quantized zero-bias conductance peak (ZBCP) of $e^2/h$ or $2e^2/h$ as a measured signature of MZMs [34], [36].

## III. Results and Discussion

### A. Model validation on the 1.6 μm-long InSb TSNW

First, we validate the Hamiltonian and the NEGF approach for Andreev processes on a proximitized 1.6 μm-long InSb NW in a magnetic field, which was numerically investigated in [18]. Figure 2 shows the normalized eigenvalue ($E/\Delta$) spectrum depending on the normalized

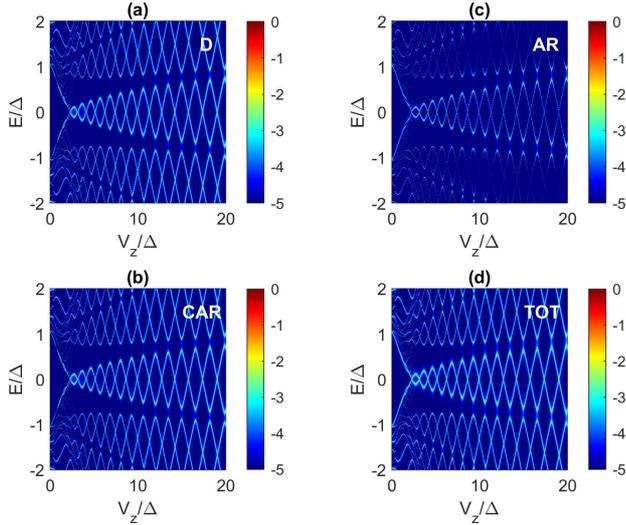

Fig. 3. Zeeman-field-dependent spectra of (a) direct, (b) CAR, (c) AR, and (d) total transmission for InSb TSNW with $L = 1.6$ μm and $\mu/\Delta = 2$.

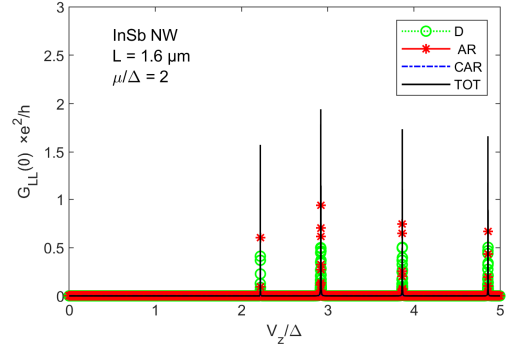

Fig. 4. Zero-bias conductance of the 1.6 μm-long InSb TSNW at $\mu/\Delta = 2$. The plot contains contributions from the direct, AR and CAR processes, and the total conductance at 0 K.

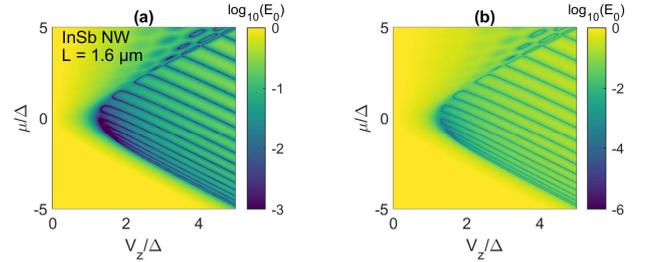

Fig. 5. Lowest-eigenenergy spectrum of the 1.6 μm-long InSb TSNW. The eigenenergies are plotted in the logarithmic scale with two different colorbar minima, and the dark shaded areas represent biasing-regions for which MZMs are expected to occur.

Zeeman field ($V_z/\Delta$) when $\mu/\Delta = 2$, i.e. for $\mu = 0.5$ meV. The spectrum is in perfect agreement with the results in [18], thus confirming our methodology. For $V_z/\Delta > 2.2$, the InSb TSNW enters the topological phase, however, MZMs appear only at very specific Zeeman field values due to the strong oscillatory behavior. The oscillations indicate that the TSNW is not sufficiently long to support MZMs with well-separated Majorana wavefunctions. Remaining in the case of $\mu/\Delta = 2$ and 1.6 μm-long InSb TSNW, we calculate the transmission functions for the direct tunneling, CAR and AR mechanisms and the total transmission and plot the respective data in Fig. 3. Direct and CAR transmissions exhibit similar properties, while the AR process is relatively weak and contributes mainly at parity crossing points, which is not easily visible in Fig. 3.

The simulated transmission enables the calculation of conductance, such that at 0 K the conductance equals the transmission function value multiplied by the conductance quantum $e^2/h$. For the Fermi level set to zero, we obtain the zero-bias local conductance for the left contact, $G_{LL}(0)$, which is plotted in Fig. 4. Sharp ZBCPs for very specific $V_z$ values are observed in Fig. 4, which indicates difficult accessibility of Majorana states for $\mu = 0.5$ meV in the InSb TSNW with $L = 1.6$ μm. The lowest two ZBCPs appear at $V_z/\Delta$ of 2.22 and 2.92, which corresponds to $B$ of 0.48 T and 0.63 T, respectively. Zhe ZBCPs should reach $2e^2/h$ due to the unitary AR transmission [6], [24], however, in Fig. 4 the conductance is somewhat lower. This difference is due to $V_z$ spectrum being insufficiently dense to pinpoint the exact $V_z$ for the parity crossing point, despite the 10001 points defined in the range $0 \leq V_z/\Delta \leq 5$. In summary, the assumed bias conditions in [18] are inadequate to guarantee feasible MZMs in the 1.6 μm-long InSb hybrid device.

### B. Zero-eigenenergy phase space

In order to facilitate finding proper external parameters, we propose to plot the lowest eigenenergy ($E_0$) phase spaces that depend on the electrochemical potential and the Zeeman or the magnetic field. Figure 5 plots the $\log_{10}(E_0/\Delta)$ in the logarithmic scale with colorbar minima going down to −3 (Fig. 5a) and −6 (Fig. 5b). The dark shaded areas indicate regions where MZMs are likely to occur. The two limits in Fig. 5a and b are chosen to indicate the design space for two different coherence time values of approximately ~1 ns and ~1 μs when $E_0 < 10^{-3}\Delta = 0.25$ μeV (Fig. 5a) and $E_0 < 10^{-6}\Delta = 0.25$ neV (Fig. 5b), respectively. The topological space with MZMs is limited by a parabola, with some approximately oval regions of zero energy modes outside of it for larger $\mu$ values. In addition, we can clearly observe multiple "Majorana lines" that are approximately parallel to the lower branch of the parabola. In comparison to the simple Kitaev chain, the zero-eigenenergy phase space of InSb TSNWs exhibits some similarities, although "Majorana lines" originate from the same point $(\mu, t) = (0, 0)$ in finite Kitaev chains [14], [37].

Figure 5 shows that for a constant $\mu/\Delta = 2$, the 1.6 μm-long InSb NW will exhibit oscillatory MZMs with very sharp peaks, as shown earlier in Figs. 2-4. This presents a practical difficulty for experiments and device reliability because it would be unlikely to pinpoint the exact $V_z$ ($B$) value for feasible and stable MZMs. On the other hand, Fig. 5a reveals a dark shaded area where $E_0 \leq 10^{-3}\Delta = 0.25$ μeV, in the vicinity of $V_z/\Delta \sim 1.56$ (~0.4 meV), equivalent to ~0.34 T, and for a negative electrochemical potential of $\mu/\Delta \sim -1$ (−0.25 meV). This narrow region presents the optimum choice of external biasing parameters ($\mu$, $V_z$ or $B$) for the InSb NW with $L = 1.6$ μm. The good news is that such design space exists, however, it is expected that optimum ranges will depend on the strength of the superconducting proximity effect and on the length of the hybrid device.

### C. Analysis of the 2.2 μm-long InSb TSNW

In this subsection we focus on the 2.2 μm-long InSb topological device that was fabricated and investigated in [5]. The exact values of material and device parameters for the EMA model are not known for these experimental hybrid devices, nevertheless, it is widely accepted that the superconducting parameter equals ~0.2-0.3 meV in the proximitized hybrid nanowires [6], [10], [15]. In the following discussions, we are interested in proving that MZMs can indeed be detected in the 2.2 μm-long InSb TSNW. Furthermore, we focus on finding the optimum design space or appropriate biasing conditions ($\mu$, $V_z$ or $B$) for Majorana modes in devices from [5].

Figure 6 shows the $\log_{10}(E_0/\Delta)$ values in the logarithmic scale for the 2.2 μm-long InSb TSNW, with the same colorbar minima as in Fig. 5. Similarly to the shorter device, the phase space of the longer TSNW is also limited by a parabola. However, the longer hybrid device exhibits more parallel "Majorana lines", and most importantly shows a considerably larger continuous MZM area than the InSb NW with $L = 1.6$ μm. The dark shaded continuous MZM area in Fig. 6a is limited by the parabola and by a straight line $(\mu/\Delta) = -3.12(V_z/\Delta) + 4.88$, with the latter enabling us to determine the limits on $\mu$ and $V_z$ ($B$). The widest $V_z$-range for potential MZMs equals ~$0.6\Delta$ (0.15 meV, equivalent to 0.13 T) and is obtained for the electrochemical potential of $-0.45\Delta$ ($-0.113$ meV).

For $\mu/\Delta = -0.45$ set by the main gate in the 2.2 μm-long InSb TSNW, we plot the zero-bias local conductance in Fig. 7. As predicted by Fig. 6, the high-$G_{LL}(0)$ region is considerably wider for $L = 2.2$ μm than for $L = 1.6$ μm. The $G_{LL}(0)$ is higher than $e^2/h$ in the ~$0.3\Delta$-wide window of the Zeeman field, which stands in contrast to a sharp ZBCP observed in Fig. 4 for the shorter device. A clear signature of MZMs, i.e. a ZBCP reaching exactly $2e^2/h$, in this $V_z$ window is reported only for $V_z/\Delta = 1.51$ (~0.38 meV), which is equivalent to $B \sim 0.33$ T. The next exactly quantized ZBCP occurs at $V_z/\Delta = 1.78$ (~0.45 meV), or equivalently at $B \sim 0.38$ T, and this conductance peak is a sharp Lorentzian similar to those observed in Fig. 4 for the shorter, 1.6 μm-long, InSb device. From the results reported in Figs. 6 and 7, we conclude that the devices reported in [5] could have hosted MZMs in the ~$0.3\Delta$-wide (~65 mT) $V_z$ ($B$) window centered around $V_z/\Delta = 1.3$ ($B \sim 0.28$ T). Indeed, when $G_{LL}(0)$ is plotted versus the magnetic field, smeared MZMs occur in an area centered at $B \sim 0.25$ T, as shown in Fig. 2 in [5], coinciding with our results reported in Fig. 7.

The positive conclusions about the 2.2 μm-long InSb hybrid devices are further strengthened by the results provided in Fig. 8 that plots the normalized LDOS for the zero-energy Majorana states. The two curves for $V_z/\Delta = 1.51$ and 1.78 correspond to the two cases of ZBCPs reported and discussed for Fig. 7. When $V_z/\Delta = 1.51$, the Majorana wavefunctions are localized close to the edges with minimum oscillations and a Majorana localization length of ~340 nm. Majorana separation equals ~1900 nm, i.e. 86% of the TSNW length. According to [38], such well separated MZMs should be feasible for TQC braiding operations. In the case when $V_z/\Delta = 1.78$, i.e. further away from the topological transition point defined by the previously mentioned parabolic limit, Majorana modes are again localized near the edges, but with significantly more oscillations and a longer localization length. The increased oscillations are a signature of stronger Majorana hybridization and a decreased coherence time [14], [39], which makes this specific bias point less feasible for practical MZMs.

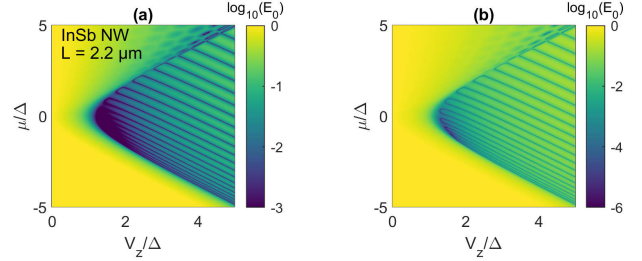

Fig. 6. Lowest-eigenenergy spectrum of the 2.2 μm-long InSb TSNW.

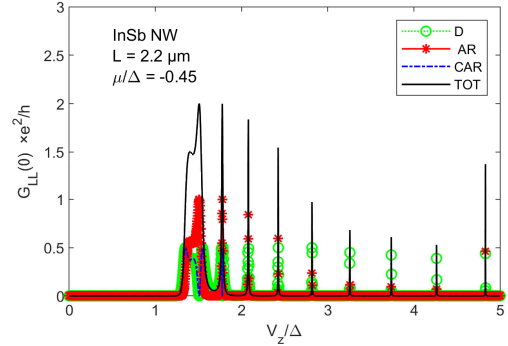

Fig. 7. Zero-bias conductance of the 2.2 μm-long InSb TSNW for $\mu/\Delta = -0.45$. The plot contains contributions from the direct, AR and CAR processes, and the total conductance at 0 K.

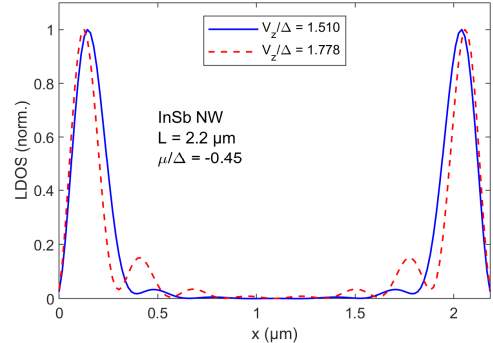

Fig. 8. Normalized LDOS in the InSb NW with $L = 2.2$ μm and for the case $\mu/\Delta = -0.45$ ($\mu = -0.113$ meV).

IV. CONLUSIONS

Tight-binding effective-mass Hamiltonians and NEGF quantum transport simulations are employed for the analysis of InSb TSNWs of various lengths in terms of their feasibility for hosting MZMs. We found that calculating spectra of the lowest eigenenergies in the biasing "phase space" ($\mu$, $V_z$ or $B$) allows us to identify optimum design space for TSNWs where MZMs are likely to occur. In the case of 2.2 μm-long InSb TSNWs, this optimum region is constrained by the parabolic limit of the topological phase and by the line $(\mu/\Delta) = -3.12(V_z/\Delta) + 4.88$, which determines the limits on appropriate $V_z$ ($B$) for the given electrochemical potential. For this specific device, we found well separated Majorana states when $\mu/\Delta = -0.45$ ($\mu = -0.113$ meV) and $V_z/\Delta = 1.51$ ($V_z = 0.38$ meV or $B = 0.33$ T). The $V_z$ window hosting high-$G_{LL}(0)$ is centered around $B \sim 0.28$ T in our numerical work and at ~0.25 T in experiments from the literature, indicating the adequacy of our methodology and the likely confirmation for the observance of MZMs in the fabricated InSb TSNWs with the length of ~2.2 μm.

ACKNOWLEDGMENTS

This work was supported by the Croatian Science Foundation under the project CONAN2D (Grant No. UIP-2019-04-3493).